\documentclass[journal]{IEEEtran}
\ifCLASSINFOpdf

\else

\fi
\usepackage{mathrsfs}%flower

\usepackage{amsmath}
\usepackage{ntheorem}

\usepackage{makeidx}  % allows for indexgeneration
\usepackage{algorithm}
\usepackage{algorithmic}
\usepackage{graphicx}
\usepackage{subfigure}
\usepackage{epstopdf}
\usepackage{bm}
\usepackage{cite}
\usepackage{stfloats}

\usepackage{amssymb}
\usepackage{graphicx}
\newtheorem{theorem}{Theorem}
\usepackage{url}

\usepackage{tabularx,booktabs}
\newcolumntype{C}{>{\centering\arraybackslash}X} % centered version of "X" type
\usepackage{lipsum}

\usepackage{makecell} % Xhline Xcline

\usepackage{graphicx}
\usepackage{color}

\newtheorem{remark}{Remark}
\newtheorem{corollary}{Corollary}

\begin{document}

\title {Statistical CSI-based Design for Reconfigurable  Intelligent Surface-aided Massive MIMO Systems with Direct Links}
\author{Kangda Zhi, Cunhua Pan, Hong Ren and Kezhi Wang\thanks{(Corresponding author: Cunhua Pan).
		 		
		K. Zhi, C. Pan are with the School of Electronic Engineering and Computer Science at Queen Mary University of London, London E1 4NS, U.K. (e-mail: k.zhi, c.pan@qmul.ac.uk).
		
		H. Ren is with the National Mobile Communications Research Laboratory, Southeast University, Nanjing 210096, China. (hren@seu.edu.cn).

K. Wang is with Department of Computer and Information Sciences, Northumbria University, UK. (e-mail: kezhi.wang@northumbria.ac.uk).}
\vspace{-10pt}
}

\maketitle
\begin{abstract}
This paper investigates the performance of reconfigurable intelligent surface (RIS)-aided massive multiple-input multiple-output (mMIMO) systems with direct links, and the phase shifts of the RIS are designed based on the statistical channel state information (CSI). We first derive the closed-form expression of the uplink ergodic data rate. Then, based on the derived expression, we use the genetic algorithm (GA) to solve the sum data rate maximization problem. With low-complexity maximal-ratio combination (MRC) and low-overhead statistical CSI-based scheme, we validate that the RIS can bring significant performance gains to the traditional mMIMO systems.
\end{abstract}

\begin{IEEEkeywords}
Intelligent reflecting surface (IRS), Reconfigurable Intelligent Surface (RIS), massive MIMO, statistical CSI.
\end{IEEEkeywords}

\vspace{-20pt}

\IEEEpeerreviewmaketitle

\section{Introduction}
Reconfigurable intelligent surface (RIS), {also known as intelligent reflecting surface (IRS)}, has been recognized as a promising technology for future wireless communication systems \cite{di2020smart,wu2020survey}. RIS is a programmable metasurface with low power consumption, and it can provide passive beamforming gains. Recently, attractive benefits of RIS have been validated in various scenarios\cite{wu2019intelligent,pan2020multicell,pan2020intelligent,boulogeorgos2020performance,yildirim2020modeling}. To be more specific, the authors in \cite{wu2019intelligent} showed that RIS has sufficient potential in future communications by jointly designing its passive beamforming and base stations (BSs)' active beamforming vectors. {The benefits of RIS for multicell MIMO communications and simultaneous wireless information and power transfer (SWIPT)  have been demonstrated in \cite{pan2020multicell} and \cite{pan2020intelligent}, respectively. From the perspective of performance analysis, the authors in \cite{boulogeorgos2020performance} proved the superiority of RIS by presenting a theoretical framework for the performance comparison of RIS and relaying systems. The authors in \cite{yildirim2020modeling} further investigated the indoor and outdoor RIS-aided systems with Rician channels.}

Different from the above instantaneous channel state information (CSI)-based contributions, extensive research attention has been shifted to exploit statistical CSI to design the phase shifts of RIS\cite{jia2020analysis,han2019large,hu2020location,Nadeem2020asym,you2020reconfigurable,zhao2020twoTime,zhi2020power}, since it can greatly reduce the channel estimation overhead and computational complexity. Specifically, { RIS-aided single-user systems with and without interference were studied in work \cite{jia2020analysis} and \cite{han2019large}, respectively. The authors in \cite{hu2020location} provided a location-based RIS design and captured the impacts of user location uncertainty. Adopting the correlated Rayleigh model, the asymptotic minimum signal-to-interference-plus-noise ratio (SINR) and energy efficiency optimization were investigated in work \cite{Nadeem2020asym} and \cite{you2020reconfigurable}, respectively. Besides, the authors in \cite{zhao2020twoTime} provided a novel two-timescale beamforming optimization scheme in RIS-aided multi-user systems. Most recently, in \cite{zhi2020power}, we firstly investigated the RIS-aided massive multiple-input multiple-output (mMIMO) systems with statistical CSI.
}

However, in\cite{zhi2020power}, only the special case was studied when the direct links from BS to users are entirely blocked. For mMIMO systems deployed in the sub-6 GHz band, the direct link may exist with high probability. Therefore, it is meaningful to consider the RIS-aided mMIMO systems with the existence of direct links. The extension to this new scenario is not straightforward. On one hand, with massive antennas, the received power from direct links could be strong, which may weaken the influence of RIS. On the other hand, with an excessive number of antennas, it is challenging to adopt complex beamforming. Instead, the maximal-ratio combination (MRC) receiver is widely used in mMIMO systems due to its simplicity and low complexity. However, due to the shared RIS-BS link between different users, massive antennas could no longer make the multi-user interference negligible. Therefore, it is imperative to investigate whether or not the RIS would still be beneficial in enhancing the performance of mMIMO systems. 

{ Different from \cite{zhi2020power}, this work further models the impacts of direct links in RIS-aided mMIMO systems, and theoretically compares the performance between IRS-aided mMIMO systems and traditional mMIMO systems. Meanwhile, the introduction of direct links also increases the complexity of the analysis. To tackle this problem, we firstly utilize the independence between cascaded links and direct links and then derive the closed-form expression for the uplink ergodic data rate. Based on this expression, we prove that even with a simple MRC receiver, RIS-aided mMIMO systems could still outperform traditional mMIMO systems. Then, a genetic algorithm (GA) is adopted to design the phase shifts of RIS which only relies on statistical CSI. Finally, simulations are carried out to verify our analytical results. }

%Against the above background, an RIS-aided massive MIMO system with direct links is studied in this paper. To reduce the implementation complexity, the MRC receiver is applied at the BS and statistical CSI is used for the design of phase shifts of the RIS. We firstly derive the closed-form expression for the uplink ergodic data rate. Then, based on the derived expression, a genetic algorithm (GA) {\color{red} is} adopted to optimize the phase shift to maximize the sum ergodic data rate. Finally, simulation results validate that statistical CSI-based RIS is capable of enhancing the performance of conventional massive MIMO systems even with simple MRC receiver.

\section{System Model}
\begin{figure}
	\setlength{\abovecaptionskip}{0pt}
	\setlength{\belowcaptionskip}{-20pt}
	\centering
	\includegraphics[width= 0.33\textwidth]{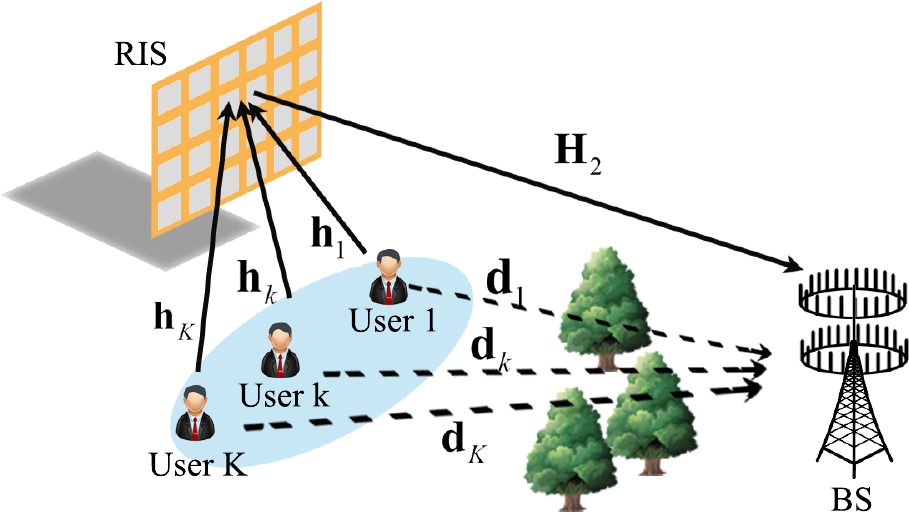}
	\DeclareGraphicsExtensions.
	\caption{An RIS-aided mMIMO system with direct links.}
	\label{figure1}
	\vspace{-10pt}
\end{figure}
As shown in Fig. \ref{figure1}, a typical uplink mMIMO system is considered, where a BS equipped with $M$ antennas simultaneously communicates with $K$ single-antenna users with the aid of an RIS. The RIS is composed of $N$ reflecting elements, and its configuration matrix can be expressed as
$ {\bf \Phi } = {\rm{diag}}\left\{ {{e^{j{\theta _1}}},{e^{j{\theta _2}}},...,{e^{j{\theta _N}}}} \right\}$ with the unit reflection amplitude,
where $\theta_n$ is the phase shift of element $n$. Then, the cascaded $M\times K$ channels can be written as ${\bf G}={\bf H}_2{\bf\Phi} {{\bf H}_1}$, where $\mathbf{H}_1=[\mathbf{h}_1,...,\mathbf{h}_K]$ is the $N\times K$ channels between users and the RIS, and $\mathbf{H}_2$ represents the $M\times N$ RIS-BS channels. 

{ Thanks to the thin and light property of RIS, there exist extensive optional deployment positions for the RIS. It can be installed on the facades of a tall building with a few scatters. Therefore, it is highly possible that LoS components exist in RIS-related channels. As in work \cite{yildirim2020modeling,han2019large,jia2020analysis,hu2020location}, we adopt the Rician fading model to respectively express the user $k$-RIS link $\mathbf{h}_k$ and RIS-BS link $\mathbf{H}_2$ as}
%Since the BS and RIS are often deployed with certain height and the RIS is often deployed near the users, channels ${\bf H}_1$ and ${\bf H}_2$ can be line-of-sight (LoS) dominant. Thus, the Rician channel model is adopted to characterize user $k$-RIS link $\mathbf{h}_k$ and RIS-BS link $\mathbf{H}_2$ respectively as
%\begin{align}
%&{{\bf h}_k} = \sqrt {{\alpha _k}} \left( {\sqrt {\frac{{{\varepsilon _k}}}{{{\varepsilon _k} + 1}}} {{\bf  \bar h}_k} + \sqrt {\frac{1}{{{\varepsilon _k} + 1}}} {{\bf \tilde h}_k}} \right),1\leq k\leq K,\\\label{rician2}
%&{{\bf H}_2} = \sqrt \beta  \left( {\sqrt {\frac{{{\delta _{}}}}{{{\delta _{}} + 1}}} {{\bf \bar H}_2} + \sqrt {\frac{1}{{\delta  + 1}}} {{\bf \tilde H}_2}} \right),
%\end{align} 
\begin{align}
&\mathbf{h}_{k}=\sqrt{\alpha_{k}}\left(\sqrt{\varepsilon_{k} /\left(\varepsilon_{k}+1\right)} \bar{\mathbf{h}}_{k}+\sqrt{1 /\left(\varepsilon_{k}+1\right)} \tilde{\mathbf{h}}_{k}\right), \\
&\mathbf{H}_{2}=\sqrt{\beta}\left(\sqrt{\delta /(\delta+1)} \bar{\mathbf{H}}_{2}+\sqrt{1 /(\delta+1)} \tilde{\mathbf{H}}_{2}\right),
\end{align} 
where $1\leq k \leq K$, $\alpha_k$ and $\beta$ are distance-dependent large-scale path-loss factors, $\varepsilon_k$ and $\delta$ are Rician factors. $\mathbf{\tilde{h}}_k \in \mathbb{C}^{N\times 1}$ and $\mathbf{\tilde{H}}_2 \in \mathbb{C}^{M\times N}$ are {non}-line-of-sight (NLoS) channel components whose elements are independent and identical distribution (i.i.d.) random variables following $\mathcal{CN}\left(0,1\right)$. By contrast, $\mathbf{\bar{h}}_k$ and $\mathbf{\bar{H}}_2$ are deterministic LoS channel components. Under the uniform square planar array (USPA) model, $\mathbf{\bar{h}}_k$ and $\mathbf{\bar{H}}_2$ can be respectively expressed as $ {{\bf \bar h}_k} = {{\bf a}_N}\left( {\varphi _{kr}^a,\varphi _{kr}^e} \right), 1\leq k \leq K, $ and $ {{\bf \bar H}_2} = {{\bf a}_M}\left( {\phi _r^a,\phi _r^e} \right){\bf a}_N^H\left( {\varphi _t^a,\varphi _t^e} \right)$ with
%\begin{align}\label{upa}
%{{\bf a}_X}\left( {\vartheta _{}^a,\vartheta _{}^e} \right) = \left[ 1,...,{e^{j2\pi \frac{d}{\lambda }\left( {x\sin \vartheta _{}^e\sin \vartheta _{}^a + y\sin \vartheta _{}^e\cos \vartheta _{}^a} \right)}},\right.\nonumber\\
%\left. ...,{e^{j2\pi \frac{d}{\lambda }\left( {\left( {\sqrt X  - 1} \right)\sin \vartheta _{}^e\sin \vartheta _{}^a + \left( {\sqrt X  - 1} \right)\sin \vartheta _{}^e\cos \vartheta _{}^a} \right)}} \right]^T,
%\end{align}
\begin{align}\label{upa}
{{\bf a}_X}\left( {\vartheta _{}^a,\vartheta _{}^e} \right) = \left[ 1,...,{e^{j2\pi \frac{d}{\lambda }\left( {x\sin \vartheta _{}^e\sin \vartheta _{}^a + y\cos \vartheta _{}^e} \right)}},\right.\nonumber\\
\left. ...,{e^{j2\pi \frac{d}{\lambda }\left( {\left( {\sqrt X  - 1} \right)\sin \vartheta _{}^e\sin \vartheta _{}^a + \left( {\sqrt X  - 1} \right)\cos \vartheta _{}^e} \right)}} \right]^T,
\end{align}
where $d$ is the elements spacing, $\lambda$ is wavelength, $\varphi _{kr}^a$, $\varphi _{kr}^e$ ($ \phi _r^a$, $\phi _r^e $) are respectively the azimuth and elevation angles of arrival (AoA) from user $k$ to the RIS (from the RIS to the BS), $ \varphi _t^a$, $\varphi _t^e $ are respectively the azimuth and elevation angles of departure (AoD) from the RIS to the BS. 

%{\color{red}As in \cite{han2019large,jia2020analysis}, we assume that these angles are perfectly known} since they can be computed from the locations of BS, RIS and users, and their locations can be obtained from the global position system (GPS).

Since rich scatters often exist near the ground, we use Rayleigh fading model to express the direct links between the BS and users { as \cite{pan2020multicell,han2019large,jia2020analysis}}. The channel of direct links $\mathbf{D} \in \mathbb{C}^{M \times K}$ can be written as
$ {\mathbf{D}=\left[\mathbf{d}_{1},\mathbf{d}_{2} ,... , \mathbf{d}_{K}\right]}, \mathbf{d}_{k}=\sqrt{\gamma_{k}} \tilde{\mathbf{d}}_{k}$,
where $\gamma_{k}$ is large-scale path loss and $\tilde{\mathbf{d}}_{k}$ represents the NLoS direct link for user $k$.

Based on the above definitions, we can express the received signal at the BS as
\begin{align}
{\bf y} =  (\mathbf{G}+\mathbf{D}) {\mathbf{P}} \mathbf{x}+\mathbf{n}=\left(\mathbf{H}_{2} {\bf\Phi} \mathbf{H}_{1}+\mathbf{D}\right) {\bf P} \mathbf{x}+\mathbf{n},   
\end{align}
where $\mathbf{P}={\rm diag}(\sqrt{p_1},...,\sqrt{p_K})$ and $p_k$ is transmit power of user $k$, $\mathbf{x}=[x_1,...,x_K]^T$ denotes the information symbol vector, $\mathbf{n} \sim \mathcal{CN}(0,\sigma^2\mathbf{I}_N)$ is the receiver noise vector.

To reduce the implementation complexity in practical systems, we employ the low-complexity MRC technique. The BS decides the MRC matrix by estimating the overall instantaneous channel matrix ${\bf G} + {\bf D}$. At the beginning of each coherence interval, users will transmit the orthogonal pilot sequences to the BS, and then the BS uses the received pilot signals to estimate the overall $M\times K$ channel $\mathbf{G}+\mathbf{D}$ by using the same method in traditional mMIMO systems\cite{Jensen2014}. We assume this channel is perfectly estimated as in \cite{han2019large,jia2020analysis} which serves as an upper bound for practical systems, and the extension to imperfect CSI scenario is left for our further work. Thus, using its perfect estimates, the  BS can configure the MRC matrix as $ (\mathbf{G}+\mathbf{D})^H $, and process the received signal as $ {\bf r} =  (\mathbf{G}+\mathbf{D})^H {\bf y} =(\mathbf{G}+\mathbf{D})^H ((\mathbf{G}+\mathbf{D}) {\mathbf{P}} \mathbf{x}+\mathbf{n})$.

{Then, the received signal at the BS that corresponds to the $k$-th user can be written as}
\begin{align}\label{signal_user_k}
&r_{k}=\sqrt{p_{k}}\left(\mathbf{g}_{k}^{H}+\mathbf{d}_{k}^{H}\right)\left(\mathbf{g}_{k}+\mathbf{d}_{k}\right) x_{k}\nonumber\\
&+\! \sum_{i=1, i \neq k}^{K} \!\!\sqrt{p_{i}}\left(\mathbf{g}_{k}^{H}+\mathbf{d}_{k}^{H}\right)\left(\mathbf{g}_{i}+\mathbf{d}_{i}\right) x_{i}+\left(\mathbf{g}_{k}^{H}+\mathbf{d}_{k}^{H}\right) \mathbf{n},
\end{align}
where $ \mathbf{g}_{k} \triangleq \mathbf{H}_{2} {\bf\Phi} \mathbf{h}_{k}$ is the $k$-th column of $\bf G$ and $\mathbf{g}_{k}$ denotes the cascaded channel of user $k$. Based on (\ref{signal_user_k}), we can express the ergodic rate expression of user $k$ as $R_{k}=\mathbb{E}\left\{  \log _{2}\left(1\!+\mathrm{SINR}_k \right)\right\}$, and the SINR of user $k$ is given by
\begin{align}\label{rate_expression}
 \mathrm{SNIR}_k \!=\!  \frac{p_{k}\left\|\mathbf{g}_{k}+\mathbf{d}_{k}\right\|^{4}}{\sum_{i=1, i \neq k}^{K} p_{i}\left|\left(\mathbf{g}_{k}^{H}+\mathbf{d}_{k}^{H}\right)\left(\mathbf{g}_{i}+\mathbf{d}_{i}\right)\right|^{2}+\sigma^{2}\left\|\mathbf{g}_{k}+\mathbf{d}_{k}\right\|^{2} }.
\end{align}

\section{Ergodic rate analysis}
%{\color{red} In this section, we  will derive the closed-form expression of the ergodic rate $R_k$.}
% Nogte that BS will user this ergodic rate to design the phase shifts of thI RIS, and then contril the RIS through the dedicated baakhaul. Therefore, this design only need to be done in a long time slot, which could significantly reduce the complexity and overhead.

%In the following Theorem, a closed-form approximation of $R_k$ is obtained.
\begin{theorem}\label{theorem1}
	In the RIS-aided mMIMO systems with the existence of direct links, the uplink ergodic data rate of user $k$ can be approximated as
\begin{align}\label{rate}
R_{k} \approx \log _{2} \! \left( \! 1 \!+  \frac{p_k E_{k}^{(signal)}({\bf\Phi})}{ \sum_{i=1, i \neq k}^{K} p_i I_{k i}({\bf\Phi})+\sigma^{2} E_{k}^{(noise)}({\bf\Phi})}  \right)\!,
\end{align}
where $E_{k}^{(signal)}({\bf\Phi})$, $I_{k i}({\bf\Phi})$ and $E_{k}^{(noise)}({\bf\Phi})$ are respectively given by (\ref{E_k_signal}), (\ref{I_ki_interference}) and (\ref{E_k_noise}) on the top of next page. 
%\newcounter{TempEqCnt}
%\setcounter{TempEqCnt}{\value{equation}} 
%\setcounter{equation}{12} 
\begin{figure*}[ht] % {\bf\Phi}
	\begin{align}\label{E_k_signal}
E_{k}^{(signal)}({\bf\Phi})&=M^{2} c_{k}^{2} \delta^{2} \varepsilon_{k}^{2}\left|f_{k}({\bf\Phi})\right|^{4}+2 c_{k} M \delta \varepsilon_{k}\left|f_{k}({\bf\Phi})\right|^{2}\left(c_{k}\left(2 M N \delta+M N \varepsilon_{k}+M N+2 M+N \varepsilon_{k}+N+2\right)+\gamma_{k}\left(M+1\right)\right) \nonumber\\
&+c_{k}^{2} M^{2} N^{2}\left(2 \delta^{2}+\varepsilon_{k}^{2}+2 \delta \varepsilon_{k}+2 \delta+2 \varepsilon_{k}+1\right)+c_{k}^{2} M N^{2}\left(\varepsilon_{k}^{2}+2 \delta \varepsilon_{k}+2 \delta+2 \varepsilon_{k}+1\right) \nonumber\\
&+c_{k} M N(M+1)\left(c_{k}\left(2 \delta+2 \varepsilon_{k}+1\right)+2 \gamma_{k}\left(\delta+\varepsilon_{k}+1\right)\right)+\gamma_{k}^{2}\left(M^{2}+M\right),   \\\label{I_ki_interference}
I_{k i}({\bf\Phi})&=M^{2} c_{k} c_{i} \delta^{2} \varepsilon_{k} \varepsilon_{i}\left|f_{k}({\bf\Phi})\right|^{2}\left|f_{i}({\bf\Phi})\right|^{2}+M c_{k} \delta \varepsilon_{k}\left|f_{k}({\bf\Phi})\right|^{2}\left(c_{i}\left(\delta M N+N \varepsilon_{i}+N+2 M\right)+\gamma_{i}\right) \nonumber\\
&+M c_{i} \delta \varepsilon_{i}\left|f_{i}({\bf\Phi})\right|^{2}\left(c_{k}\left(\delta M N+N \varepsilon_{k}+N+2 M\right)+\gamma_{k}\right)+M N^{2} c_{k} c_{i}\left(M \delta^{2}+\delta\left(\varepsilon_{i}+\varepsilon_{k}+2\right)+\left(\varepsilon_{k}+1\right)\left(\varepsilon_{i}+1\right)\right) \nonumber\\
&+M^{2} N c_{k} c_{i}\left(2 \delta+\varepsilon_{i}+\varepsilon_{k}+1\right)+M^{2} c_{k} c_{i} \varepsilon_{k} \varepsilon_{i}\left|\overline{\mathbf{h}}_{k}^{H} \overline{\mathbf{h}}_{i}\right|^{2}+2 M^{2} c_{k} c_{i} \delta \varepsilon_{k} \varepsilon_{i} \operatorname{Re}\left\{f_{k}^{H}({\bf\Phi}) f_{i}({\bf\Phi}) \overline{\mathbf{h}}_{i}^{H} \overline{\mathbf{h}}_{k}\right\} \nonumber\\
&+M\left(c_{i} \gamma_{k} N\left(\delta+\varepsilon_{i}+1\right)+c_{k} \gamma_{i} N\left(\delta+\varepsilon_{k}+1\right)+\gamma_{i} \gamma_{k}\right),\\\label{E_k_noise}
E_{k}^{(noise)}({\bf\Phi})&=M\left(c_{k} \delta \varepsilon_{k}\left|f_{k}({\bf\Phi})\right|^{2}+c_{k}\left(\delta+\varepsilon_{k}+1\right) N+\gamma_{k}\right).
	\end{align}
	%\hrulefill
\end{figure*}
Besides, $c_{k}\triangleq \frac{\beta \alpha_{k}}{(\delta+1)\left(\varepsilon_{k}+1\right)}$, $f_{k}({\bf\Phi}) \triangleq \mathbf{a}_{N}^{H}\left(\varphi_{t}^{a}, \varphi_{t}^{e}\right) {\bf\Phi} \overline{\mathbf{h}}_{k}=\sum_{n=1}^{N} e^{j\left(\zeta_{n}^{k}+\theta_{n}\right)}$, where 
\begin{align}
\zeta_{n}^{k}=&2 \pi \frac{d}{\lambda}  \left(   \lfloor(n-1) / \sqrt{N}\rfloor  \left(\sin \varphi_{kr}^{e} \sin \varphi_{kr}^{a}-\sin \varphi_{t}^{e} \sin \varphi_{t}^{a}\right)\right.\nonumber\\
&\left.+((n-1) \bmod \sqrt{N})\left( \cos \varphi_{kr}^{e}- \cos \varphi_{t}^{e}\right)\right).
\end{align}
\end{theorem} 
%\begin{align}
%\begin{array}{l}
%\zeta_{n}^{k}=2 \pi \frac{d}{\lambda}  \left(   \lfloor(n-1) / \sqrt{N}\rfloor  \left(\sin \varphi_{kr}^{e} \sin \varphi_{kr}^{a}-\sin \varphi_{t}^{e} \sin \varphi_{t}^{a}\right)\right.\\
%\qquad\left.+((n-1) \bmod \sqrt{N})\left( \cos \varphi_{kr}^{e}- \cos \varphi_{t}^{e}\right)\right).
%\end{array}
%\end{align}

\itshape {Proof:}  \upshape Please refer to Appendix A. \hfill $\blacksquare$

Theorem \ref{theorem1} shows that expression (\ref{rate}) does not depend on the fast-varying instantaneous CSI $\tilde{\mathbf{h}}_k$, $\tilde{\mathbf{H}}_2$ and $\tilde{\mathbf{d}}_k$, but it only relies on the statistical CSI, i.e., the AoA and AoD in $\bar{\mathbf{h}}_k$ and $\bar{\mathbf{H}}_2$, the path-loss coefficients $\beta,\alpha_k, \gamma_{k}$ and Rician factors $\delta,\varepsilon_{k}$, which change very slowly. Specifically, environment-related path-loss coefficients and Rician factors can be measured and saved at the BS in advance. The BS can also estimate the AoA and AoD based on global position systems\cite{hu2020location}. Therefore, to simplify our analysis and draw key insights, we assume that the statistical CSI is perfectly known at the BS before the data transmission\cite{han2019large,jia2020analysis,Jensen2014}. Then, based on Eq. (\ref{rate}), the BS exploits statistical CSI to design and update the phase shifts of RIS. Note that this operation only needs to be done once for a long time, which significantly reduces the computational complexity and feedback overhead.

\begin{remark}
	The ergodic data rate for RIS-aided mMIMO systems without direct links in \cite{zhi2020power} can be obtained by setting $\gamma_k=0,\forall k$. Besides, the single-user case in \cite{han2019large} is a special case of our work with $p_i=0,\forall i\neq k$.
\end{remark}

\begin{corollary}\label{coro1}
	The ergodic data rate for traditional mMIMO systems without an RIS can be obtained by setting $c_k=0,\forall k$, which is given by  $R_{k}^\mathrm{(w)}\triangleq \log _{2}\left(1+{\rm SINR}_{k}^\mathrm{(w)}\right)$ with $ {\rm SINR}_{k}^\mathrm{(w)}  \approx        {p_k (M+1)\gamma_k    }  \left/  {\left(  { \sum_{i=1, i \neq k}^{K} p_i   \gamma_{i}   +\sigma^{2}    }\right) } \right.$.
	
	%$ {\color{red}{\rm SINR}_{k}^\mathrm{(w)}  \approx   \frac{p_k (M+1)\gamma_k    }{ \sum_{i=1, i \neq k}^{K} p_i   \gamma_{i}   +\sigma^{2}    }} $

\end{corollary}

Corollary \ref{coro1} shows that under a simple MRC receiver but with a large $M$, the power of interference is negligible compared with the signal. However, this feature no longer holds for RIS-aided mMIMO systems. We can find that both the signal term in (\ref{E_k_signal}) and interference term in (\ref{I_ki_interference}) are on the order of $\mathcal{O}(M^2)$, which means that a large $M$ will make the signal-interference-ratio converge to a constant. Thus, a basic question is that whether this additional interference will limit the gain of RIS. To answer this question and provide clear insights, we will use some special cases to compare RIS-aided mMIMO systems with non-RIS-aided mMIMO systems.

\begin{corollary}
	For a special case where cascaded channels are pure NLoS, i.e., $\delta=\varepsilon_{k}=0,\forall k$, the ergodic rate is $R_{k}^\mathrm{(NL)}    \triangleq  \log _{2}   \left(  1  + \mathrm{SINR}_{k}^\mathrm{(N L)}  \right)   $ and $\mathrm{SINR}_{k}^\mathrm{(N L)}$ is given by (\ref{rate_Nlos}) on the top of this page.
	\begin{figure*}
		\vspace{-25pt}
	\begin{align}\label{rate_Nlos}
        \mathrm{SINR}_{k}^\mathrm{(N L)}  \approx    \frac{p_{k}\left(c_{k}^{2}\left(M N^{2}+N^{2}+MN+N\right)+2 c_{k} \gamma_{k} N \left(M+1\right)+\gamma_{k}^{2}(M+1)\right)}{\sum_{i=1, i \neq k}^{K} p_{i}\left(c_{k} c_{i}\left(N^{2}+M N\right)+c_{k} \gamma_{i} N+c_{i} \gamma_{k} N+\gamma_{k} \gamma_{i}\right)+\sigma^{2}\left(c_{k} N+\gamma_{k}\right)}. \quad\qquad\qquad\qquad\qquad\qquad
\end{align}

\hrulefill
	\end{figure*}
\end{corollary}

It can be seen that in this special case, there is no need to design the phase shifts of RIS. Besides, when both $N\to\infty$ and $M\to\infty$, rate $R_{k}^\mathrm{(NL)}$ could grow without bound. Furthermore, \cite[Fig. 4]{zhi2020power} has shown that this rich-scatter environment is favorable for RIS-aided muti-user systems since it can provide sufficient spatial multiplexing gains. Therefore, we will use this special case to investigate the gain from RIS in the presence of additional interference. To further facilitate our analysis and provide useful insights, we focus on the two-user case where users are located closely and have the same transmit power, i.e., $c_k=c,\gamma_{k}=\gamma,p_k=p, k=1,2$. Then,  by solving the inequality  $\mathrm{SINR}_{k}^\mathrm{(N L)} > \mathrm{SINR}_{k}^\mathrm{(w)} $, we can obtain the following Corollary.
\begin{corollary}\label{corollary4}
Considering an RIS deployed in the environment with rich scatters and $\gamma>0,c> 0$, the ergodic data rate of RIS-aided mMIMO systems is higher than non-RIS-aided mMIMO systems when
\begin{align}
\frac{p}{\sigma^{2}}\!<\!\frac{N+1}{\gamma(M\!-\!1)}\!+\!\frac{1}{c(M\!-\!1)}, \text{ or } {N}\!>\!  \gamma \!\left(  \!  \frac{p}{\sigma^{2}}(M-1)  -  \frac{1}{c}  \right)  -1    .
\end{align}
\end{corollary}
%\begin{align}
%\frac{p}{\sigma^{2}}<\frac{N+1}{\gamma(M-1)}+\frac{1}{c(M-1)}, \text{ or } {N}>  \gamma \left(    \frac{p}{\sigma^{2}}(M-1)  -  \frac{1}{c}  \right)  -1    .
%\end{align}
%\itshape {Proof:}  \upshape , we can complete the proof after some simplifications. \hfill $\blacksquare$

Corollary \ref{corollary4} indicates that RIS-aided mMIMO can outperform non-RIS-aided systems under three cases: 1) in low-SNR regime; 2) with large $N$; 3) with weak direct links strength $\gamma$. Reasons behind these results are: 1) In high SNR regime, rate will be interference-limited which aggravates the negative impacts of RIS's additional interference; 2)  large $N$ can increase the passive beamforming gain of RIS; 3) with strong direct links, the signal contributions from cascaded links become relatively small.

Next, we study another case where the phase shifts of the RIS are adjusted randomly. For analytical tractability, we focus on an extreme scenario where $N\to\infty$. This case is reasonable since RIS is comprised of low-cost passive elements, and large $N$  can also help RIS unleash its passive beamforming gains.
\begin{corollary}
 Assume that RIS's phase shifts are set randomly in each time block. When $N\to\infty$, the average data rate is given by $R_{k}^\mathrm{(r m)} \triangleq \log _{2}\left(1+\mathrm{SINR}_{k}^\mathrm{(r m)}\right) $  with
\begin{align}
\mathrm{SINR}_{k}^\mathrm{(r m) } \approx  \frac{p_{k} \alpha_{k}\left(M\left(2 \delta^{2}+2 \delta+1\right)+2 \delta+1\right)}{\sum_{i=1, i \neq k}^{K} p_{i} \alpha_{i}\left(M \delta^{2}+2 \delta+1\right)}.
\end{align}
\end{corollary}
\itshape {Proof:}  \upshape Using the same method as \cite[Corollary 4]{zhi2020power}, we can substitute terms involving $\bf\Phi$ in (\ref{rate}) with their expectation. When $N\to\infty$, we can ignore the insignificant terms which are not on the order of $\mathcal{O}(N^2)$ and then complete the proof after some simplifications. \hfill $\blacksquare$

Then, we consider a two-user case with $ c_k=c,\gamma_{k}=\gamma,\varepsilon_{k}=\varepsilon, p_k=p, k=1,2 $. By solving  inequality $\mathrm{SINR}_{k}^\mathrm{(rm)} > \mathrm{SINR}_{k}^\mathrm{(w)} $, we can obtain the following result.
\begin{corollary}\label{corollary6}
	When $N\to\infty$, RIS-aided mMIMO systems with random phase shifts outperform traditional non-RIS systems if
	\begin{align}\label{inequality1}
\gamma	\frac{p}{\sigma^{2}}<\frac{\left(2 \delta^{2}+2 \delta+1\right)M+(2 \delta+1)}{\delta^{2}\left(M^{2}-M\right)}.
	\end{align}
\end{corollary}

Corollary \ref{corollary6} shows that when RIS's phase shifts are set randomly, to beat non-RIS systems, it should operate in low-SNR regime. Besides, even when $N\to\infty$, it is still challenging to meet inequality (\ref{inequality1}) under large $M$.

\section{Numerical Results}
In this section, numerical simulations are presented to verify the correctness of our analytical results. Based on Eq. (\ref{rate}), a GA-based method is utilized to design the optimal phase shifts of RIS, where its framework is shown in Algorithm \ref{algorithm1}.
	\begin{algorithm}[t]
	\caption{GA-based Method}
	\begin{algorithmic}[1]\label{algorithm1}
		\STATE Initialize a population of 200 individuals where individual $t$ has a randomly generated chromosome $\mathbf{\Phi}_t$;  Count = 1;
		\WHILE{ Count $\leq 100*N$}
		\STATE Calculate the fitness of each individual as $R(t)=\sum_{k=1}^{K} R_k(\mathbf{\Phi}_t)$, where $R_k(\bf\Phi)$ is given in (\ref{rate});
		\STATE Remove the top 10 individuals with higher fitness from the current population as elites;
		\STATE Remove 40 individuals with lower fitness from the current population, and use uniform mutation\cite{zhi2020power} with probability 0.1 to create 40 offspring;
		\STATE Generate 300 parents from remaining individuals based on stochastic universal sampling\cite{zhi2020power}, and then perform two-points crossover\cite{zhi2020power} to create 150 offspring;
		\STATE Combine 10 elites, 40 offspring and 150 offspring to evolve to the next-generation; Count = Count+1;
		\ENDWHILE
		\STATE Output the chromosome of the individual with the highest fitness in the current population.

	\end{algorithmic}
\end{algorithm}
%We use the same GA-based method in \cite{zhi2020power} to maximize the sum user rate $R=\sum_{k=1}^{K} R_k$.

%\begin{figure}
%	\setlength{\abovecaptionskip}{0pt}
%	\setlength{\belowcaptionskip}{-20pt}
%	\centering
%	\includegraphics[width= 0.25\textwidth]{fig2.pdf}
%	\DeclareGraphicsExtensions.
%	\caption{The top view of our simulation setup.}
%	\label{figure2}
%	\vspace{-10pt}
%\end{figure}
%Our simulation setup is shown in Fig. \ref{figure2}. 

Unless stated otherwise, we set $M=N=49, \sigma^2=-104\text{ dBm}, p_k=30\text{ dBm},\varepsilon_{k}=10,\forall k, \delta=1$. Four users are evenly located on a circle centered at the RIS with a radius of $d_\mathrm{UI}=20$ m as in \cite[Fig. 6]{wu2019intelligent}. RIS-BS distance is $d_\mathrm{IB}=1000$ m and the distance between user $k$ and BS is calculated by $(d_k^\mathrm{UB})^2={\left(d_\mathrm{IB}-d_\mathrm{UI} \sin \left(\frac{\pi}{5} k\right)\right)^{2}+\left(d_\mathrm{UI} \cos \left(\frac{\pi}{5} k\right)\right)^{2}}$. All the AoA and AoD are generated randomly from $[0,2\pi]$\cite{pan2020multicell,pan2020intelligent}. The distance-based path-loss are $\alpha_{k}=10^{-3}d_\mathrm{UI}^{-2}, \beta=10^{-3}d_\mathrm{IB}^{-2.5}$ and $\gamma_{k}=10^{-3}\left(d_{k}^\mathrm{UB}\right)^{-4},\forall k$.

\begin{figure*}
	\setlength{\abovecaptionskip}{-5pt}
	\setlength{\belowcaptionskip}{-15pt}
	\centering
	\begin{minipage}[t]{0.33\linewidth}
		\centering
		\includegraphics[width=2.3in]{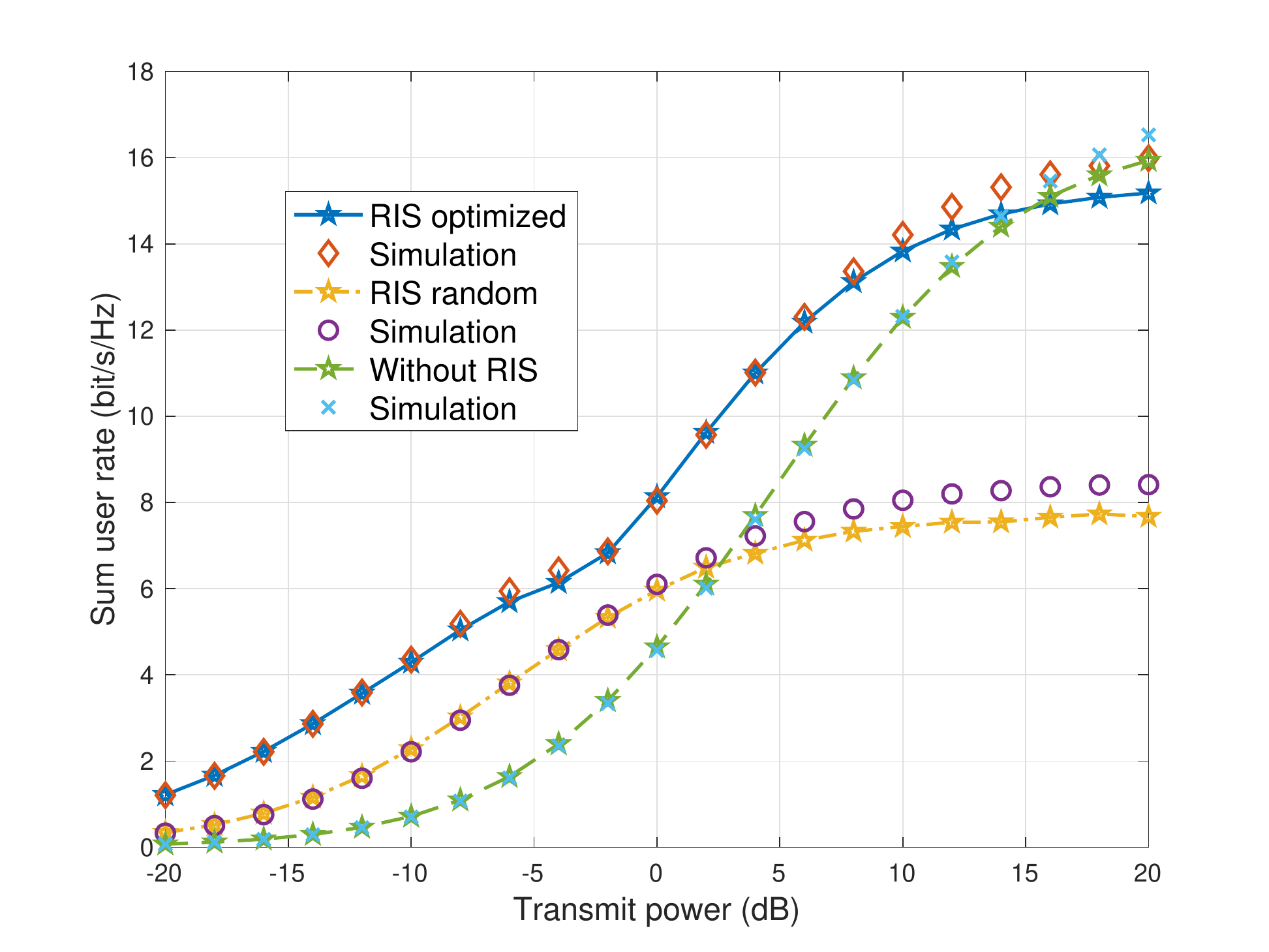}
		\caption{Rate versus transmit power.}
		\label{figure3}
	\end{minipage}%
	\begin{minipage}[t]{0.33\linewidth}
		\centering
		\includegraphics[width=2.3in]{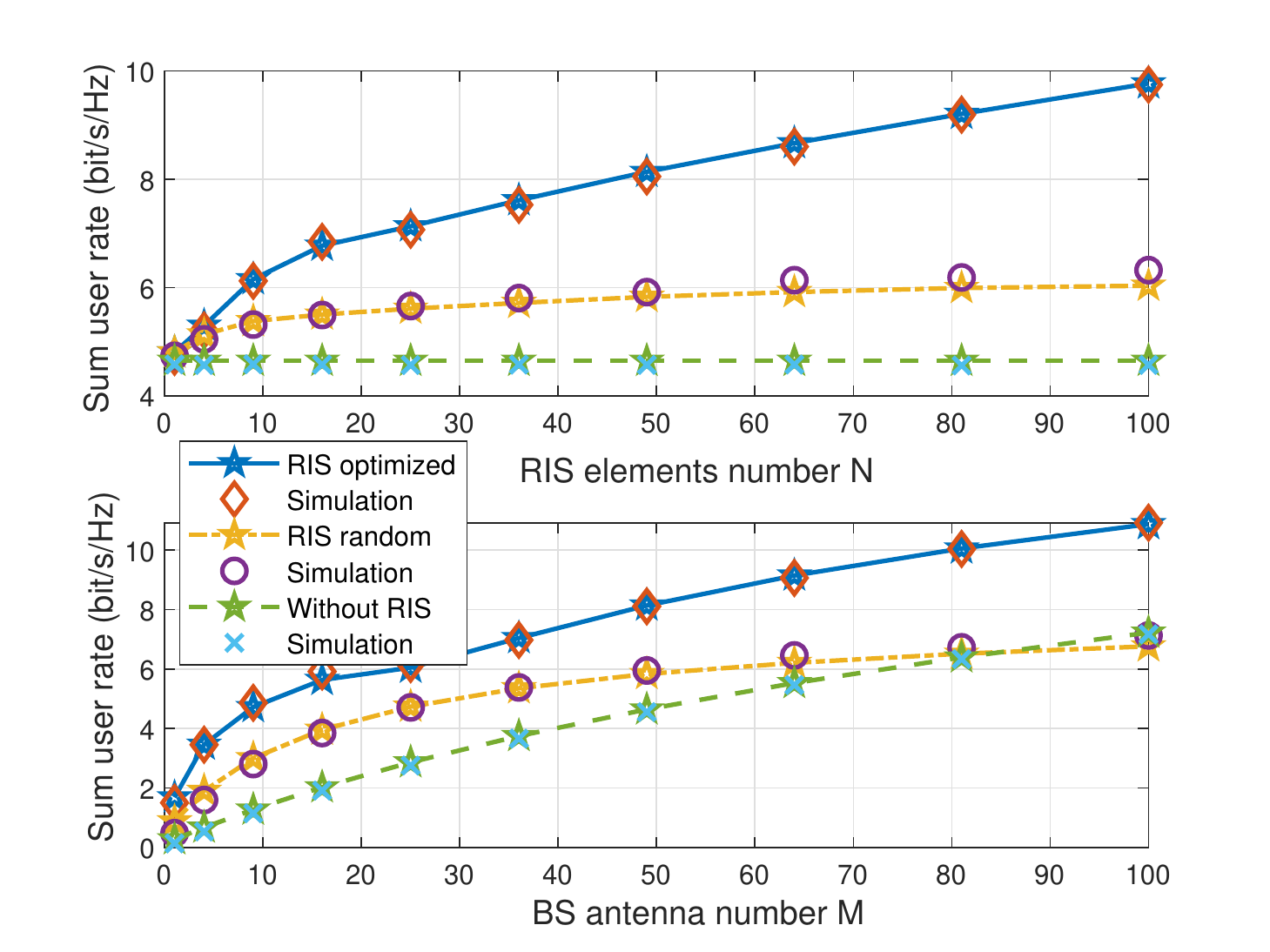}
		\caption{Rate versus $ M $ and $N$ with $d_\mathrm{IB}=1000$ m.}
		\label{figure4}
	\end{minipage}
	\begin{minipage}[t]{0.33\linewidth}
		\centering
		\includegraphics[width=2.3in]{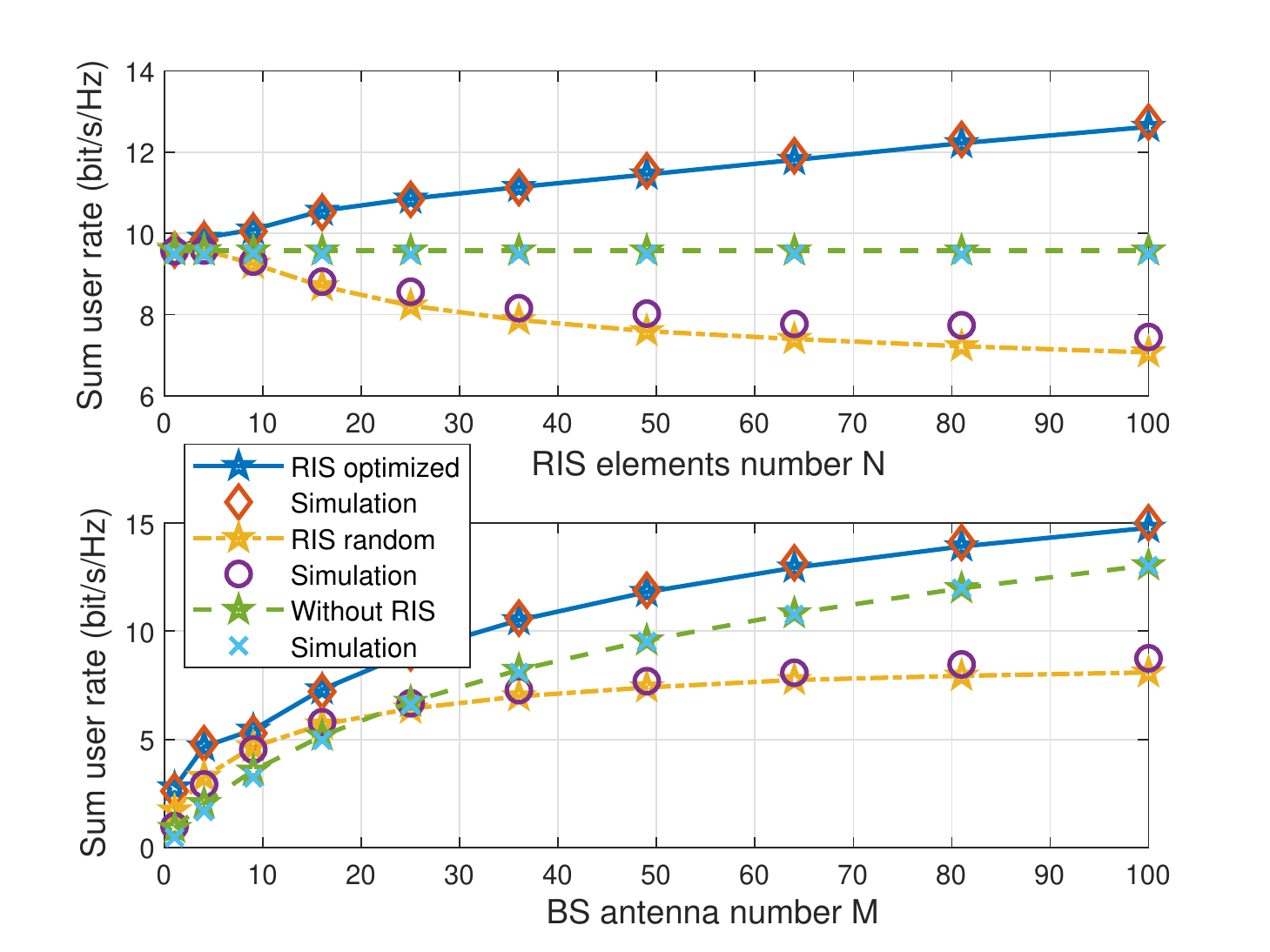}
		\caption{Rate versus $ M $ and $N$ with $d_\mathrm{IB}=700$ m.}
		\label{figure5}
	\end{minipage}
	\vspace{-10pt}
\end{figure*}

%\begin{figure}
%	\setlength{\abovecaptionskip}{0pt}
%	\setlength{\belowcaptionskip}{-20pt}
%	\centering
%	\includegraphics[width= 0.4\textwidth]{fig3.eps}
%	\DeclareGraphicsExtensions.
%	\caption{Uplink ergodic rate versus user transmit power.}
%	\label{figure3}
%	\vspace{-10pt}
%\end{figure}
%
%\begin{figure}[t]
%	\setlength{\abovecaptionskip}{0pt}
%	\setlength{\belowcaptionskip}{-20pt}
%	\centering
%	\includegraphics[width= 0.40\textwidth]{fig4.eps}
%	\DeclareGraphicsExtensions.
%	\caption{Rate versus $ M $ and $N$ with $d_\mathrm{IB}=1000$.}
%	\label{figure4}
%	\vspace{-10pt}
%\end{figure}
%
%\begin{figure}
%	\setlength{\abovecaptionskip}{0pt}
%	\setlength{\belowcaptionskip}{-20pt}
%	\centering
%	\includegraphics[width= 0.40\textwidth]{fig5.eps}
%	\DeclareGraphicsExtensions.
%	\caption{Rate versus $ M $ and $N$ with $d_\mathrm{IB}=700$.}
%	\label{figure5}
%	\vspace{-10pt}
%\end{figure}

Fig. \ref{figure3} shows that even with a simple MRC receiver, statistical CSI-based RIS can still effectively improve the rate performance in mMIMO systems in the low-SNR regime. However, due to the multi-user interference, as SNR increases, conventional mMIMO systems will outperform random phase shifts-based RIS systems. Finally, with extremely high SNR, it could even outperform the optimal phase shifts-based RIS systems. These results agree with our analysis in Corollary \ref{corollary4} and Corollary \ref{corollary6}.

Fig. \ref{figure4} and Fig. \ref{figure5} plot the ergodic rate versus $N$ and $M$ with $d_\mathrm{IB}=1000$ {m} and $d_\mathrm{IB}=700$ {m}, respectively. Note that smaller $d_\mathrm{IB}$ means stronger direct links. We can see that the RIS with optimal phase shifts brings a significant rate improvement to traditional mMIMO systems in both figures, and this improvement can hold with quite large $M$. However, when the direct links are strong or when $M$ is large, mMIMO systems without RIS have a better performance than RIS-aided systems with random phase shifts. This result is consistent with our analysis in Corollary \ref{corollary6}. Besides, these observations indicate that to fully take advantages of RIS with a simple MRC receiver, it is better to use RIS with a large number of elements to serve cell-edge users, and the RIS has ability to play a significant role in the low-SNR regime.

%if low-complexity MRC beamforming is applied, the phase shifts of IRS can be set randomly for users in the cell-edge. However, to serve users not too far away from the BS, the phase shifts of IRS should be carefully designed.
\vspace{-10pt}
\section{Conclusion}
In this paper, we have studied an RIS-aided mMIMO system with direct links. A closed-form ergodic rate expression has been derived. Then, based on the expression, we have found that with low-complexity MRC beamforming, RIS-aided mMIMO systems can outperform non-RIS systems in the low-SNR regime. Finally, our analytical results have been verified by the simulations.

\vspace{-20pt}
\begin{appendices}
	
\section{}
To begin with, by applying \cite[Lemma 1]{Jensen2014}, ergodic data rate $R_{k}=\mathbb{E}\left\{  \log _{2}\left(1\!+\mathrm{SINR}_k \right)\right\}$ can be approximated as
\begin{align}\label{rate_appro}
\begin{array}{l}
R_{k}\!\approx \! \log _{2}\!\!\left( \!\!1 + \! \frac{p_{k} \mathbb{E}\left\{ \left\|\mathbf{g}_{k}+\mathbf{d}_{k}\right\|^{4}   \right\}   }     { \sum_{i=1, i \neq k}^{K} \!p_{i}    \mathbb{E}\left\{\left|\left(\mathbf{g}_{k}^{H}+\mathbf{d}_{k}^{H}\right)\left(\mathbf{g}_{i}+\mathbf{d}_{i}\right)\right|^{2}\right\}+\sigma^{2}\mathbb{E}\left\{\left\|\mathbf{g}_{k}+\mathbf{d}_{k}\right\|^{2}\right\} } \!\!\right)\!\!.
\end{array}
\end{align}

To derive a closed-form expression of (\ref{rate_appro}), we need to derive $\mathbb{E}\left\{ \left\|\mathbf{g}_{k}+\mathbf{d}_{k}\right\|^{4}   \right\}$,  $\mathbb{E}\left\{\left|\left(\mathbf{g}_{k}^{H}+\mathbf{d}_{k}^{H}\right)\left(\mathbf{g}_{i}+\mathbf{d}_{i}\right)\right|^{2}\right\}$ and  $\mathbb{E}\left\{\left\|\mathbf{g}_{k}+\mathbf{d}_{k}\right\|^{2}\right\}$, respectively. Note that $\mathbf{d}_{k}$ are independent to $\mathbf{g}_{k}$ and $\mathbf{d}_{i},\forall i\neq k$, and $\mathbf{d}_{k}$ is composed of i.i.d. entries with zero mean, we can firstly derive the noise term as
\begin{align}\label{noise_term}
\begin{array}{l}
\mathbb{E}\left\{\left\|\mathbf{g}_{k}+\mathbf{d}_{k}\right\|^{2}\right\}=\mathbb{E}\left\{\left(\mathbf{g}_{k}+\mathbf{d}_{k}\right)^{H}\left(\mathbf{g}_{k}+\mathbf{d}_{k}\right)\right\} \\
=\mathbb{E}\left\{\mathbf{g}_{k}^{H} \mathbf{g}_{k}+\mathbf{g}_{k}^{H} \mathbf{d}_{k}+\mathbf{d}_{k}^{H} \mathbf{g}_{k}+\mathbf{d}_{k}^{H} \mathbf{d}_{k}\right\} \\
=\mathbb{E}\left\{\mathbf{g}_{k}^{H} \mathbf{g}_{k}+\mathbf{d}_{k}^{H} \mathbf{d}_{k}\right\}=\mathbb{E}\left\{\left\|\mathbf{g}_{k}\right\|^{2}\right\}+\mathbb{E}\left\{\left\|\mathbf{d}_{k}\right\|^{2}\right\} \\
=\mathbb{E}\left\{\left\|\mathbf{g}_{k}\right\|^{2}\right\}+\gamma_{k} M,
\end{array}
\end{align}
where $\mathbb{E}\left\{\left\|\mathbf{g}_{k}\right\|^{2}\right\}$ has been given in \cite[Lemma 1]{zhi2020power}.

Next, the signal term $\mathbb{E}\left\{ \left\|\mathbf{g}_{k}+\mathbf{d}_{k}\right\|^{4}   \right\}$ can be expanded as
\begin{align}\label{E4}
\begin{array}{l}
\mathbb{E}\left\{\left\|\mathbf{g}_{k}+\mathbf{d}_{k}\right\|^{4}\right\} \\
=\mathbb{E}\left\{\left(\left\|\mathbf{g}_{k}\right\|^{2}+2 \operatorname{Re}\left\{\mathbf{d}_{k}^{H} \mathbf{g}_{k}\right\}+\left\|\mathbf{d}_{k}\right\|^{2}\right)^{2}\right\} \\
=\mathbb{E}\left\{\left\|\mathbf{g}_{k}\right\|^{4}\right\}+4 \mathbb{E}\left\{\left(\operatorname{Re}\left\{\mathbf{d}_{k}^{H} \mathbf{g}_{k}\right\}\right)^{2}\right\}\\
\quad+\mathbb{E}\left\{\left\|\mathbf{d}_{k}\right\|^{4}\right\}+2 \mathbb{E}\left\{\left\|\mathbf{g}_{k}\right\|^{2}\left\|\mathbf{d}_{k}\right\|^{2}\right\},
\end{array}
\end{align}
where $\mathbb{E}\left\{\left\|\mathbf{g}_{k}\right\|^{4}\right\}$ has been given in \cite[Lemma 1]{zhi2020power}. Assuming that $ {\left[\mathbf{g}_{k}\right]_{m}=v_{m}+j w_{m}}$ and ${\left[\mathbf{d}_{k}^{H}\right]_{m}=s_{m}+j t_{m}} $, where both $s_m$ and $t_m$ independently follow  $\mathcal{N}\left(0,\frac{\gamma_{k}}{2}\right)$, we have
\begin{align}\label{E4_1}
\begin{array}{l}
\mathbb{E}\left\{\left(\operatorname{Re}\left\{\mathbf{d}_{k}^{H} \mathbf{g}_{k}\right\}\right)^{2}\right\} \\
=\mathbb{E}\left\{\left(\sum_{m=1}^{M} s_{m} v_{m}-t_{m} w_{m}\right)^{2}\right\} \\
=\mathbb{E}\left\{\sum_{m=1}^{M}\left(s_{m} v_{m}-t_{m} w_{m}\right)^{2}\right\} \\
=\mathbb{E}\left\{\sum_{m=1}^{M}\left(s_{m} v_{m}\right)^{2}+\left(t_{m} w_{m}\right)^{2}\right\} \\
=\frac{\gamma_{k}}{2} \mathbb{E}\left\{\sum_{m=1}^{M}\left(v_{m}\right)^{2}+\left(w_{m}\right)^{2}\right\} \\
=\frac{\gamma_{k}}{2} \mathbb{E}\left\{\left\|\mathbf{g}_{k}\right\|^{2}\right\}.
\end{array}
\end{align}

Then, the remaining two terms in (\ref{E4}) can be obtained as
\begin{align}\label{E4_2}
\begin{array}{l}
\mathbb{E}\left\{\left\|\mathbf{d}_{k}\right\|^{4}\right\}=\mathbb{E}\left\{\left(\sum_{m=1}^{M}\left|\left[\mathbf{d}_{k}\right]_{m}\right|^{2}\right)^{2}\right\} \\
=\mathbb{E} \! \left\{\sum\limits_{m=1}^{M}\left|\left[\mathbf{d}_{k}\right]_{m}\right|^{4}\right\}  \! + \!  \mathbb{E} \!  \left\{\sum\limits_{m_1=1}^{M} \sum\limits_{m_2=1 \atop m_2 \neq m_1}^{M}  \!  \!  \left|\left[\mathbf{d}_{k}\right]_{m_1}\right|^{2}\left|\left[\mathbf{d}_{k}\right]_{m_2}\right|^{2}\right\} \\
=2 M \gamma_{k}^{2}+M(M-1)\gamma_{k}^{2} = \left(M^{2}+M\right)\gamma_{k}^{2},
\end{array}
\end{align}
and
\begin{align}\label{E4_3}
\mathbb{E}\left\{\left\|\mathbf{g}_{k}\right\|^{2} \! \left\|\mathbf{d}_{k}\right\|^{2}\right\} \! = \!  \mathbb{E}\left\{\left\|\mathbf{g}_{k}\right\|^{2}\right\}  \!  \mathbb{E}\left\{\left\|\mathbf{d}_{k}\right\|^{2}\right\} \! = \! M\gamma_{k}  \mathbb{E}\left\{\left\|\mathbf{g}_{k}\right\|^{2}\right\} \! .
\end{align}

Substituting (\ref{E4_1}), (\ref{E4_2}) and (\ref{E4_3}) into (\ref{E4}), the expression of signal term is given by
\begin{align}\label{signal_term}
\begin{array}{l}
\mathbb{E}\left\{\left\|\mathbf{g}_{k}+\mathbf{d}_{k}\right\|^{4}\right\} \\
=\mathbb{E}\left\{\left\|\mathbf{g}_{k}\right\|^{4}\right\}+2 (M+1) \gamma_{k}\mathbb{E}\left\{\left\|\mathbf{g}_{k}\right\|^{2}\right\}+\left(M^{2}+M\right)\gamma_{k}^{2}.
\end{array}
\end{align}

Finally, the interference term can be written as
%\begin{align}\label{interference_term}
%\begin{array}{l}
%\mathbb{E}\left\{\left|\left(\mathbf{g}_{k}^{H}+\mathbf{d}_{k}^{H}\right)\left(\mathbf{g}_{i}+\mathbf{d}_{i}\right)\right|^{2}\right\} \\
%=\mathbb{E}\left\{\left|\mathbf{g}_{k}^{H} \mathbf{g}_{i}+\mathbf{d}_{k}^{H} \mathbf{g}_{i}+\mathbf{g}_{k}^{H} \mathbf{d}_{i}+\mathbf{d}_{k}^{H} \mathbf{d}_{i}\right|^{2}\right\} \\
%=\mathbb{E}\left\{\left|\mathbf{g}_{k}^{H} \mathbf{g}_{i}\right|^{2}\right\}+\mathbb{E}\left\{\left|\mathbf{d}_{k}^{H} \mathbf{g}_{i}\right|^{2}\right\}+\mathbb{E}\left\{\left|\mathbf{g}_{k}^{H} \mathbf{d}_{i}\right|^{2}\right\}+\mathbb{E}\left\{\left|\mathbf{d}_{k}^{H} \mathbf{d}_{i}\right|^{2}\right\},
%\end{array}
%\end{align}
\begin{align}\label{interference_term}
\begin{array}{l}
\mathbb{E}\left\{\left|\left(\mathbf{g}_{k}^{H}+\mathbf{d}_{k}^{H}\right)\left(\mathbf{g}_{i}+\mathbf{d}_{i}\right)\right|^{2}\right\} \\
=\mathbb{E}\left\{\left|\mathbf{g}_{k}^{H} \mathbf{g}_{i}+\mathbf{d}_{k}^{H} \mathbf{g}_{i}+\mathbf{g}_{k}^{H} \mathbf{d}_{i}+\mathbf{d}_{k}^{H} \mathbf{d}_{i}\right|^{2}\right\} \\
=\mathbb{E}\! \left\{\left|\mathbf{g}_{k}^{H} \mathbf{g}_{i}\right|^{2}\! \right\}  \!+\!   \mathbb{E}\! \left\{\left|\mathbf{d}_{k}^{H} \mathbf{g}_{i}\right|^{2}\! \right\}   \!+\!   \mathbb{E}\! \left\{\left|\mathbf{g}_{k}^{H} \mathbf{d}_{i}\right|^{2}\! \right\}  \!+\!   \mathbb{E}\! \left\{\left|\mathbf{d}_{k}^{H} \mathbf{d}_{i}\right|^{2}\! \right\},
\end{array}
\end{align}
where
$\mathbb{E}\left\{\left|\mathbf{g}_{k}^{H} \mathbf{g}_{i}\right|^{2}\right\}$ has been given in \cite[Lemma 1]{zhi2020power}, and
\begin{align}
\begin{array}{l}
\mathbb{E}\left\{\left|\mathbf{d}_{k}^{H} \mathbf{g}_{i}\right|^{2}\right\} 
=\mathbb{E}\left\{\mathbf{g}_{i}^{H} \mathbb{E}\left\{\mathbf{d}_{k} \mathbf{d}_{k}^{H}\right\} \mathbf{g}_{i}\right\} 
=\gamma_{k} \mathbb{E}\left\{\left\|\mathbf{g}_{i}\right\|^{2}\right\},\\
\mathbb{E}\left\{\left|\mathbf{g}_{k}^{H} \mathbf{d}_{i}\right|^{2}\right\}
=\mathbb{E}\left\{\mathbf{g}_{k}^{H} \mathbb{E}\left\{\mathbf{d}_{i} \mathbf{d}_{i}^{H}\right\} \mathbf{g}_{k}\right\} 
=\gamma_{i} \mathbb{E}\left\{\left\|\mathbf{g}_{k}\right\|^{2}\right\},\\
\mathbb{E}\left\{\left|\mathbf{d}_{k}^{H} \mathbf{d}_{i}\right|^{2}\right\} 
=\mathbb{E}\left\{\mathbf{d}_{k}^{H} \mathbb{E}\left\{\mathbf{d}_{i} \mathbf{d}_{i}^{H}\right\} \mathbf{d}_{k}\right\} 
=\gamma_{i} \gamma_{k} M.
\end{array}
\end{align}

Therefore, combining (\ref{noise_term}), (\ref{signal_term}) and (\ref{interference_term}) with (\ref{rate_appro}) and after some simplifications, we can complete the proof. 
\end{appendices}

%%%%%%%%%%%%%%%%%%%%%%%%%%%%%%%%%%%%% Reference
\bibliographystyle{IEEEtran}
\vspace{-6pt}
\bibliography{myref.bib}
%\end{thebibliography}
\end{document}